%% file: main.tex
\documentclass[sigconf]{acmart}
\AtBeginDocument{%
  }

\copyrightyear{2026}
\acmYear{2026}
\setcopyright{cc}
\setcctype{by}
\acmConference[ISWC '26]{Proceedings of the 2026 ACM International Symposium on Wearable Computers}{October 11--15, 2026}{Shanghai, China}
\acmBooktitle{Proceedings of the 2026 ACM International Symposium on Wearable Computers (ISWC '26), October 11--15, 2026, Shanghai, China}
\acmDOI{10.1145/3830727.3834819}
\acmISBN{979-8-4007-2872-3/2026/10}

\usepackage{enumitem}
\usepackage{hyperref}
\usepackage{subcaption}
\usepackage{float}
\usepackage{graphicx}
\usepackage{subcaption}
\usepackage{xcolor}
\usepackage{soul}
\usepackage{afterpage}
\usepackage{listings}
\usepackage{array}
\usepackage{multirow}
\usepackage{bm}
\usepackage{arydshln}
\usepackage{amsmath}
\usepackage{caption}
\usepackage{stfloats}
\usepackage{balance}

\input{participant_summary.tex}
\input{questionnaire_statistics.tex}

\sethlcolor{yellow} 

\newbool{showComments}
\booltrue{showComments}
\usepackage{listings}
\usepackage{array}
\usepackage{multirow}
\booltrue{showComments}
\ifbool{showComments}{%

}

\begin{document}

\title{Finger-to-Ear ECG: Systematic Evaluation of an Earbud-Based Cardiac Monitoring Approach}

\author{Philipp Lepold}
\email{philipp.lepold@kit.edu}
\orcid{0009-0003-0391-588X}
\affiliation{%
  \institution{Karlsruhe Institute of Technology}
  \city{Karlsruhe}
  \country{Germany}
}

\author{Paula Breitling}
\email{paula.breitling@kit.edu}
\orcid{0000-0001-7242-8805}
\affiliation{%
  \institution{Karlsruhe Institute of Technology}
  \city{Karlsruhe}
  \country{Germany}
}

\author{Jonas Hummel}
\email{jonas.hummel@kit.edu}
\orcid{0009-0005-8563-6175}
\affiliation{%
  \institution{Karlsruhe Institute of Technology}
  \city{Karlsruhe}
  \country{Germany}
}

\author{Tobias Röddiger}
\email{tobias.roeddiger@ipai-foundation.ai}
\orcid{0000-0002-4718-9280}
\affiliation{%
  \institution{IPAI Foundation gGmbH}
  \city{Heilbronn}
  \country{Germany}
}

\author{Michael Beigl}
\email{michael.beigl@kit.edu}
\orcid{0000-0001-5009-2327}
\affiliation{%
  \institution{Karlsruhe Institute of Technology}
  \city{Karlsruhe}
  \country{Germany}
}

\renewcommand{\shortauthors}{Philipp Lepold, Paula Breitling, Jonas Hummel, Tobias Röddiger, and Michael Beigl}

\begin{abstract}

Ear-ECG enables unobtrusive cardiac monitoring, but existing approaches often struggle with low signal amplitudes and limited morphology preservation. We evaluate a finger-to-ear ECG paradigm that combines an in-ear electrode with a rear housing finger-contact electrode in a speaker-equipped earbud. A study with 30 participants investigated four finger-to-ear electrode geometries against chest-reference ECG and additionally assessed robustness under music playback, talking, and walking disturbances. Left-finger configurations achieved R peak detection F1-scores >99\% with high agreement of key ECG morphology features. In contrast, right finger configurations as well as walking led to severe signal degradation. The results reveal a trade-off: while cross-body setups provided the highest signal fidelity, users preferred same-side contacts for comfort reasons. This establishes finger-to-ear ECG as a robust, morphology-aware paradigm for opportunistic sensing in earables.

\end{abstract}


\begin{teaserfigure}
  \includegraphics[width=\textwidth]{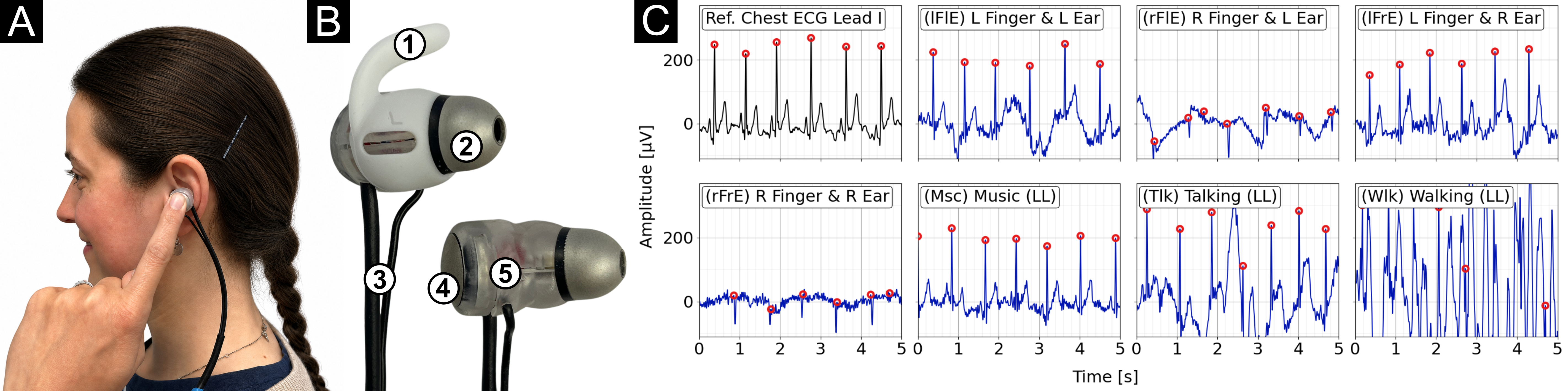}
  \caption{[A] Person performing a finger-to-ear ECG recording; the left index finger is placed on the rear earable electrode; [B]~Depiction of left and right ear-ECG earbuds: (1): Optional silicone ear-hook to improve secure fit, (2): In-ear electrode, (3): Separate wires for ECG \& music signals, reducing interference, (4): Rear finger electrode, (5): 3D printed housing; [C]: Filtered ECG recordings from one participant (5 s) of all tested configurations \& noise conditions. Red circles mark recognised R peaks.}
  \label{fig:teaser}
\end{teaserfigure}

\maketitle

\section{Introduction}
Continuous cardiac monitoring is of increasing medical and scientific interest, enabling the early detection of arrhythmias and cardiovascular disease~\cite{dahiya_wearable_2024, kamga_use_2022}. Although multi-lead electrocardiography (ECG) remains the clinical gold standard, conventional Holter systems are impractical for extended use due to body-worn electrodes and cables. Commercial wearables, like the Apple Watch, demonstrate that single-lead wearable ECG can achieve clinically meaningful performance, but also highlight the trade-off between usability and the richer spatial information provided by clinical ECG~\cite{klier_diagnostic_2023,alnasser2023reliability}.

Recently, the ear has emerged as a promising site for physiological sensing due to its proximity to relevant anatomical structures, low susceptibility to motion artefacts, and integration into everyday devices such as earbuds and hearing aids~\cite{roddiger2022sensing}. Previous work demonstrated the feasibility of head and in-ear ECG sensing, including morphology-preserving recordings under controlled conditions \cite{pullin2025ear, von_rosenberg_hearables_2017, hammour_hearables_2019,yarici_hearables_2024}. However, reliable ear-ECG acquisition remains challenging because cardiac potentials measurable at the head are comparatively weak, making signal quality highly sensitive to electrode placement, contact stability, and user interaction.

To address these limitations, we propose a finger-to-ear ECG approach that combines an in-ear electrode with a rear earbud electrode contacted by the index finger. 
We systematically evaluate all four finger-ear geometries against chest-reference ECG under stationary and realistic disturbance conditions (talking, audio playback, walking), and show that selected configurations preserve ECG morphology features within a practical earable form factor.

\section{Related Work}
Early work on ear-based ECG explored unconventional electrode placements around the ear and head to capture cardiac activity. 
\citet{shen_ear-lead_2008} demonstrated the feasibility of ear-based ECG using an additional arm electrode, while \citet{celik_evaluation_2016} investigated multiple electrode placements around the head and body. Subsequent work further showed that head- and ear-based measurements can preserve ECG morphology under controlled conditions~\cite{von_rosenberg_hearables_2017,yarici_hearables_2024}. 

Following \citet{looney2012ear}'s concept of in-ear EEG electrodes, later studies demonstrated the feasibility of in-ear ECG sensing across the ear canals~\cite{von_rosenberg_hearables_2017, hammour_hearables_2019}. However, in a comprehensive review of sensing with earables, \citet{roddiger2022sensing} note that only a few studies report acceptable ear-ECG performance, highlighting a central challenge of ear-ECG: signal quality strongly depends on electrode geometry, anatomy, and user behaviour, and stable, morphology-preserving recordings cannot yet be achieved consistently across all users and form factors. For example, \citet{mandic_your_2023} improved single-ear ECG quality by adding a mastoid electrode, but this resulted in a bulky form factor compared to commercial earbuds.

Given the inherently low signal amplitude and susceptibility to artefacts in ear-ECG, several works focus on signal enhancement and denoising~\cite{zhang_hear_2017, zhang2017machine}, with recent approaches leveraging deep learning for robust signal reconstruction and peak detection~\cite{davies_deep-match_2024, santos_real-time_2025, occhipinti2024ear}. These approaches improve R peak recovery under weak acquisition conditions, but primarily focus on heartbeat detection rather than preserving diagnostically relevant waveform morphology. Consequently, improving the acquisition geometry itself remains an open challenge for ear-ECG systems.

To address these limitations, recent research has explored alternative interaction paradigms and electrode placements. \citet{de_lucia_-ear_2021} introduced the concept of finger-to-ear ECG using the left index finger, resulting in promising signal quality under controlled conditions. 
However, their work evaluated only a single finger-to-ear configuration without investigating robustness under realistic disturbances or user acceptance of different interaction geometries. 
This leaves open whether other finger-ear combinations yield superior signal quality and if they remain reliable during common disturbances. \citet{palmisciano_eyewear-based_2025} similarly showed that an integrated finger electrode in smart glasses can produce usable ECG signals, while \citet{guler_wearable_2025} showed the broader potential of interaction-based ECG acquisition through temporary body contacts. 

However, neither work utilises the ear canal nor systematically compares acquisition geometries, leaving open how electrode placement affects signal quality, morphology preservation, and usability in ear-centred ECG systems. This motivates the systematic evaluation presented in this work.

\section{Finger-to-Ear ECG Earbuds}
The proposed finger-to-ear ECG earbuds combine an in-ear electrode (measuring electrode) with a finger electrode (reference electrode) exposed on the rear side of the earbud housing, without an additional ground (DRL) electrode. This choice was made to preserve a simple and realistic earbud design, in which only one electrode is placed in the ear canal; common-mode interference is suppressed by an instrumentation amplifier and subsequent software filtering. ECG recordings are initiated by touching the rear electrode with the index finger during wear, as shown in \autoref{fig:teaser}(A). A standard earphone driver speaker (RE6, Huayunxin) is integrated into the earbuds to enable audio playback functionality. 

\begin{figure}[h]
    \centering
    \includegraphics[width=\linewidth]{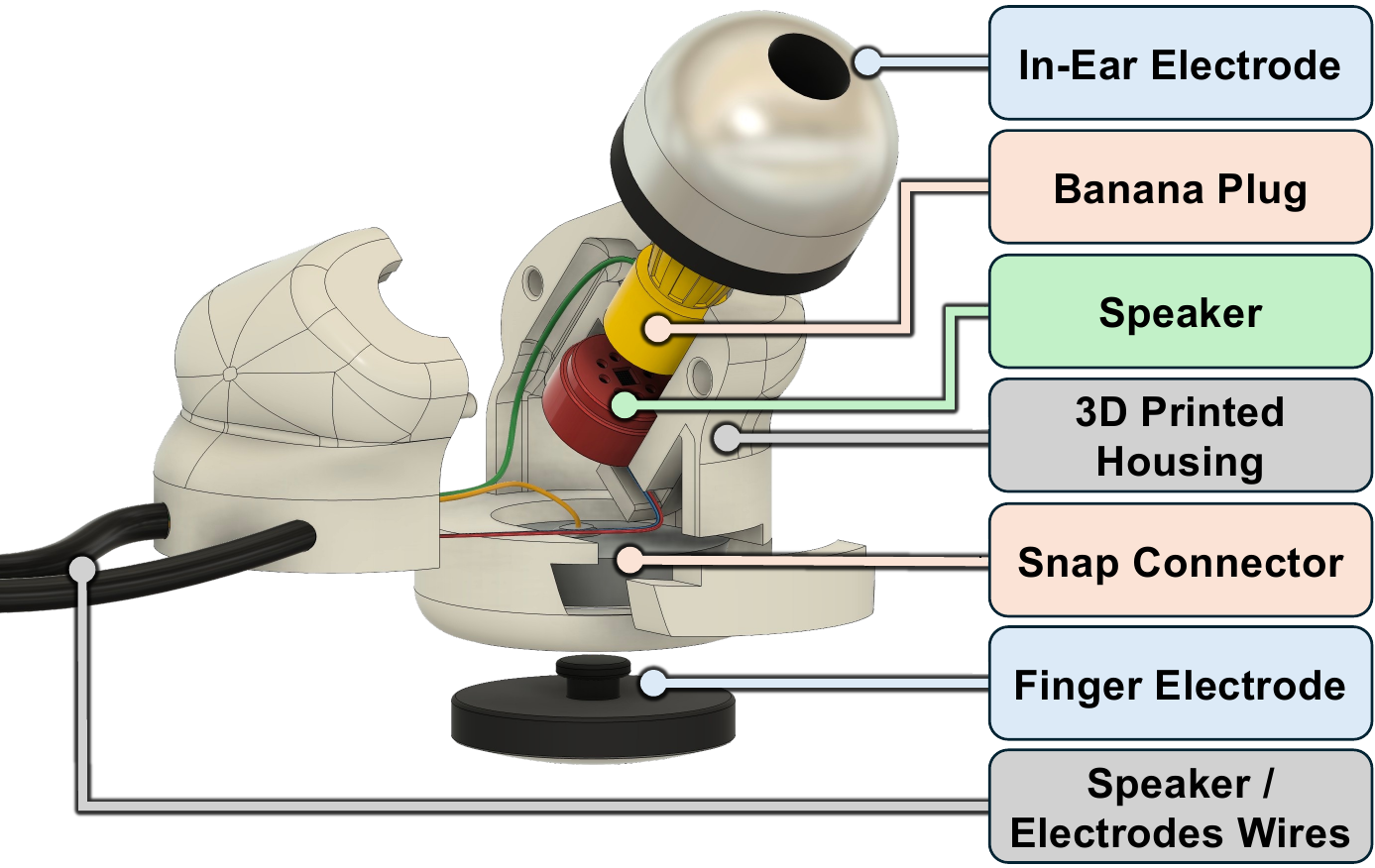}
    \caption{Exploded-view illustration of the earbud design.}
    \label{fig:earbud-explosion}
\end{figure}

\autoref{fig:teaser}(B) depicts the earbuds and \autoref{fig:earbud-explosion} shows a rendering of the system design.
The hardware design combines elements from the open-source OpenEarable\,2.0 and OpenEarable ExG platforms \cite{roddiger2025openearable, lepold2024openearable}. 
Specifically, the outer earbud shell and mechanical in-ear form factor are based on the OpenEarable\,2.0 earpiece design, while the banana-plug electrode interface originates from the OpenEarable ExG platform. 
CAD files of the design resources are provided publicly on GitHub under the MIT licence.\footnote{\url{https://github.com/teco-kit/Finger-to-Ear-ECG}} 

The earbud housing was manufactured using SLA 3D printing. In-ear electrodes are connected through a standard 3.5\,mm banana plug integrated into the housing. For the finger electrode, a commercially available ECG snap connector was mechanically modified by removing its plastic housing, leaving only the conductive metal snap stud embedded into the rear side of the earbud enclosure to interface with snap-on dome electrodes. Both electrodes are dry conductive polymer electrodes coated with silver/silver-chloride (Softpulse{\texttrademark}, Dätwyler).
Shielded cables were used for ECG connections; the speaker was driven separately through a standard 3.5\,mm headphone jack to minimise interference.

\section{Evaluation}
To provide a systematic analysis of all possible finger-to-ear combinations, including common real-world disturbances, we collected a dataset of finger-to-ear ECG with \Participants{} adults (\AgeSummary{}, \SexSummary{}, \HandednessSummary{}) and evaluated the signal using the metrics outlined in this section. The study was conducted in accordance with our university's ethical standards; all participants provided written informed consent.

\subsection{Experimental Design}  
\begin{figure}[h]
    \centering
    \includegraphics[width=\linewidth]{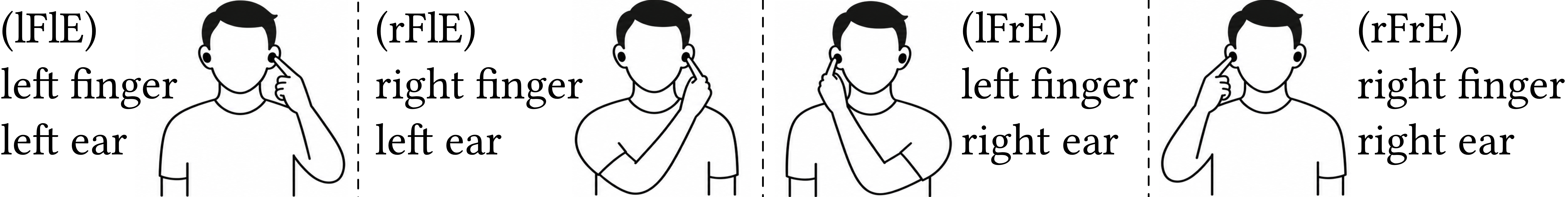}
    \caption{The four evaluated finger-to-ear electrode configurations: \textit{lFlE}, \textit{rFlE}, \textit{lFrE} and \textit{rFrE}, from left to right.}
    \label{fig:pictograms}
\end{figure}

Ear-ECG and chest-reference ECG were recorded simultaneously in two parts. First, participants completed a comparison of four finger-to-ear geometries illustrated in \autoref{fig:pictograms}: left finger \& left ear (\textit{lFlE}), right finger \& left ear (\textit{rFlE}), left finger \& right ear (\textit{lFrE}), and right finger \& right ear (\textit{rFrE}). Second, robustness under disturbances was evaluated only for \textit{lFlE}, as it is consistent with prior work \cite{de_lucia_-ear_2021}, provides a comfortable interaction, and allows disturbance effects to be assessed relative to a single baseline, as these primarily affect electrode contact rather than electrode geometry.

The three disturbance conditions tested were: listening to music through the integrated earbud speakers at 50\% volume (\textit{Msc}), reading a Wikipedia article aloud (\href{https://en.wikipedia.org/wiki/Alpine_chough}{\textit{Alpine chough}}) (\textit{Tlk}), and walking repeatedly along a corridor (\textit{Wlk}). 

For each condition, we recorded 180\,seconds of ear-ECG in randomised order, separated by 60\,s pauses between trials. Participants placed their elbow on the table and remained still during all stationary conditions. 
After each stationary condition and after walking, participants rated their agreement with the statements that \textit{"The arm position is comfortable"}, \textit{"[...] feels natural"}, and \textit{"[...] causes pain"} on a 7-point Likert scale (1 = strongly disagree, 7 = strongly agree). 

The ECG-earbuds were connected to a modified OpenEarable ExG analog frontend (AD7124-4, Analog Devices \& COSINA333, Cosine) \cite{lepold2024openearable} as an extension board connected to the main PCB of an OpenEarable 2.0 \cite{roddiger2025openearable}, recording with 256 Hz onto a microSD card.
A 3-lead ECG (Faros\texttrademark, Bittium) was used as the ground truth with silver/silver-chloride wet electrodes, sampling at 250 Hz.

\subsection{Signal Processing}
The ear-ECG signal recorded using OpenEarable was synchronised with UTC before each recording, therefore serving as the temporal reference. The ear- and reference ECG signals were notch-filtered (50\,Hz; 2\,Hz bandwidth), then bandpass filtered (0.5 - 30\,Hz), both with 4th-order Butterworth filters. The ear-ECG signal was subsequently downsampled to 250\,Hz to match the reference ECG. To compensate for temporal misalignment, the reference ECG was aligned to the ear-ECG time base using cross-correlation-based lag estimation and linear drift correction.
Segments were excluded after manual inspection only when technical failures made analysis impossible, namely flatline recordings or persistent high-amplitude noise throughout the recording due to device malfunctions. Analysts were not blinded; exclusions were based solely on objectively unusable recordings, independent of experimental condition. In total, 11 of 210 segments (5.2\%) were excluded, with no more than three exclusions per condition.
An example recording after filtering across all configurations can be seen in \autoref{fig:teaser}(C).

R peaks were detected using \textit{NeuroKit2’s} \cite{Makowski2021neurokit} ECG peak detector, and P, QRS and T wave fiducial points extracted via its DWT-based delineation algorithm.
RR intervals were extracted from each ECG recording (ear-ECG and reference ECG) as the temporal intervals between consecutive R wave peaks. Signals were compared with lead I of the chest-reference ECG, as prior work found finger-based ear-ECG configurations to show the closest correspondence to lead I morphology \cite{shen_ear-lead_2008, de_lucia_-ear_2021, palmisciano_eyewear-based_2025}. To assess whether wearable finger-to-ear ECG preserves not only heartbeat timing but also diagnostically relevant waveform morphology, we derived complementary metric families based on prior ECG and ear-ECG literature. Following established heart-rate-variability recommendations and prior wearable ECG evaluations \cite{palmisciano_eyewear-based_2025, electrophysiology1996heart, shaffer2017overview, de_lucia_-ear_2021, yarici_hearables_2024}, we included heartbeat detection as well as rhythm-based metrics derived from RR intervals to quantify temporal agreement with the reference ECG, and analysed waveform similarity together with fiducial-point-based ECG morphology features to evaluate preservation of atrial and ventricular waveform characteristics beyond heartbeat detection alone \cite{rautaharju2009aha, surawicz2009aha}. The following features were derived:

\subsubsection{Heartbeat Metrics}
Based on RR intervals, several standard heart rate and heart rate variability (HRV) indices were derived:

\begin{enumerate}[label=\roman*]

\item \textbf{Heart Rate}. Heart rate in beats per minute (bpm); $\overline{RR}$ denotes the mean RR interval per participant \& configuration:

\( HR = \frac{60}{\overline{RR}} \)

\item \textbf{Standard Deviation of NN Intervals}. The variability of normal-to-normal (NN) R intervals; a measure of HRV:

\( SDNN = \sqrt{\frac{1}{n-1}\sum_{i=1}^{n}(RR_i-\overline{RR})^2} \)

\item \textbf{Root Mean Square of Successive Differences}. Reflects short-term beat-to-beat variability:

\( RMSSD = \sqrt{\frac{1}{n-1}\sum_{i=1}^{n-1}(RR_{i+1}-RR_i)^2} \)

\item \textbf{Percentage of Successive RR Differences Exceeding 50\,ms}. Short-term variations in cardiac rhythm:

\( pNN50 = 100 \cdot \frac{\sum_i \mathbf{1}(|RR_{i+1}-RR_i| > 50\,\mathrm{ms})}{n-1} \)

\end{enumerate}

\subsubsection{Heartbeat Detection}
Performance was evaluated separately for each recording configuration and summarised across participants using the median value.

\begin{enumerate}[label=\roman*.]
    
    \item \textbf{Precision, Recall, and F1 Score:}  
    Ear-ECG heartbeats were matched against reference beats using a temporal tolerance window of $\pm 40$\,ms around the R peak, as in \citet{davies_deep-match_2024}.

    \item \textbf{Absolute Heart Rate Error} [bpm]: 
    $\Delta HR = \left| HR_{\mathrm{ear}} - HR_{\mathrm{ref}} \right|$
    
    \item \textbf{Absolute HRV Error} [ms]: 
    $\Delta HRV = \left| SDNN_{\mathrm{ear}} - SDNN_{\mathrm{ref}} \right|$
    
\end{enumerate}
	
\subsubsection{ECG Waveform Metrics}
	Metrics were derived from the fiducial points within each heartbeat; beats with missing detections were excluded. Values were independently aggregated for ear-ECG and reference ECG using participant-level median.
	
	\begin{enumerate}[label=\roman*.]
		
		\item \textbf{P Wave Amplitude (P\textsubscript{A}):}  
		Amplitude of a detected P wave peak.
        
		\item \textbf{QRS Complex Amplitude (QRS\textsubscript{A}):}  
		Peak-to-peak amplitude excursion across the detected Q, R, and S fiducial points.
		
		\item \textbf{PR Interval:}  
		Time between P wave \& QRS complex onset.
		
		\item \textbf{QRS Duration:}  
		Duration of ventricular depolarisation (onset to offset).
		
		\item \textbf{QT Interval:}  
		Time from ventricular depolarisation onset to repolarisation.
		
	\end{enumerate}

\begin{figure*}[b]
    \centering
    \includegraphics[width=\textwidth]{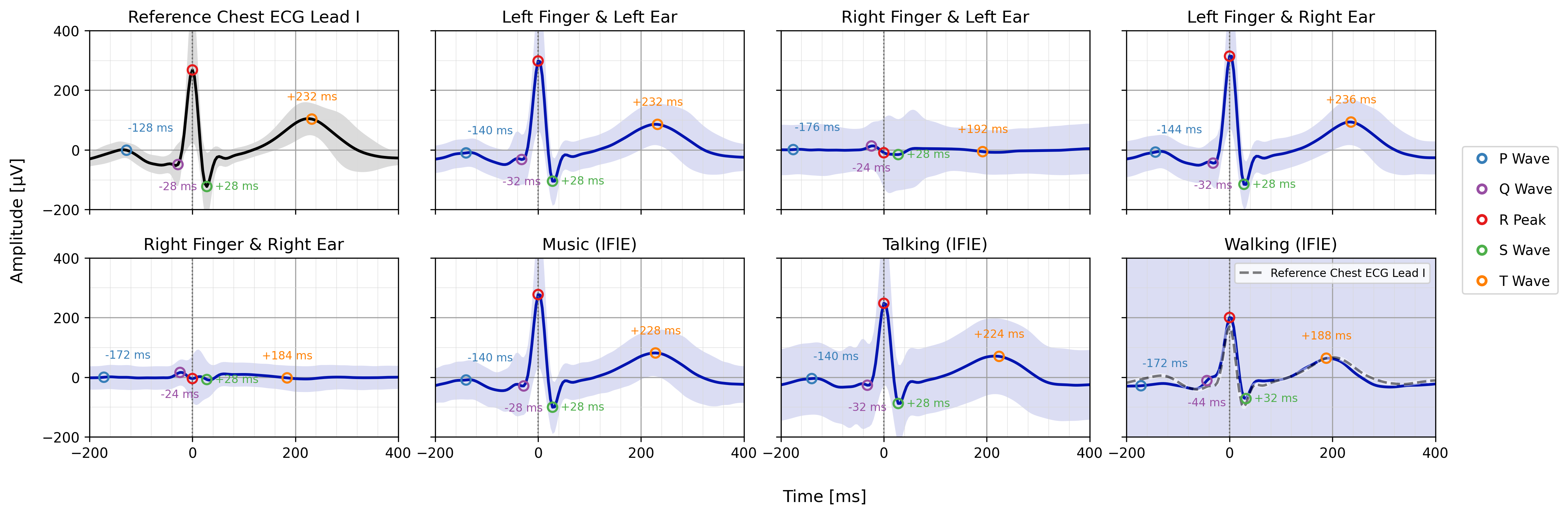}
    \caption{Average ECG waveforms (mean $\pm$ 1 standard deviation) for all configurations and noise conditions, and chest-reference. Signals are centred around all detected R peaks, then averaged within a 600 ms window (-200 ms \& +400 ms). The median timings of the P, Q, R, S and T waves are marked on the waveform as circles (from left to right).}
    \label{fig:avg-waveform}
\end{figure*}

\section{Results}
\input{resources/yarici_mean_cardiac_rhythm_correlation_values.tex}
All correlation analyses used Pearson correlation coefficients and an initial significance level of $\alpha = 0.05$. Bonferroni correction was applied separately within each analysis family: waveform correlations (7 tests, $\alpha_{\mathrm{adj}} = 0.0071$), HR/HRV metrics (28 tests, $\alpha_{\mathrm{adj}} = 0.0018$), and ECG morphology metrics (35 tests, $\alpha_{\mathrm{adj}} = 0.0014$).

\subsection{Signal Quality}
\subsubsection{Average Waveform}
\autoref{fig:avg-waveform} shows mean ECG waveforms for all configurations and reference ECG, averaged within a 600\,ms window around each detected R peak (200\,ms before and 400\,ms after), including median P, Q, R, S and T wave timings.
Left-finger configurations \textit{lFlE} ($r=\YaricilFlEMeanCardiacRhythmPearson$) and \textit{lFrE} ($r=\YaricilFrEMeanCardiacRhythmPearson$) showed the highest similarity to reference morphology with consistent PQRS fiducial timings, reflected by significant waveform correlations. Both configurations exhibited slightly earlier median P wave (12 - 16\,ms) and Q wave (4\,ms) onsets compared to the reference ECG. In contrast, the right-finger configurations showed substantially altered waveforms and negative waveform correlations (\textit{rFlE}: $r=\YaricirFlEMeanCardiacRhythmPearson$, \textit{rFrE}: $r=\YaricirFrEMeanCardiacRhythmPearson$), indicating poor morphology preservation.

Relative to \textit{lFlE}, the waveforms of the disturbance conditions \textit{Msc} and \textit{Tlk} remained visually similar, with only slightly (\textit{Msc}: $r=\YariciMscMeanCardiacRhythmPearson$) and moderately (\textit{Tlk}: $r=\YariciTlkMeanCardiacRhythmPearson$) increased waveform variability. 
In contrast, \textit{Wlk} caused substantial waveform degradation with barely distinguishable P and Q waves, despite retaining moderate overall waveform similarity ($r=\YariciWlkMeanCardiacRhythmPearson$).

\subsubsection{Heartbeat Detection}
\begin{figure}[h]
    \centering
    \includegraphics[width=\linewidth]{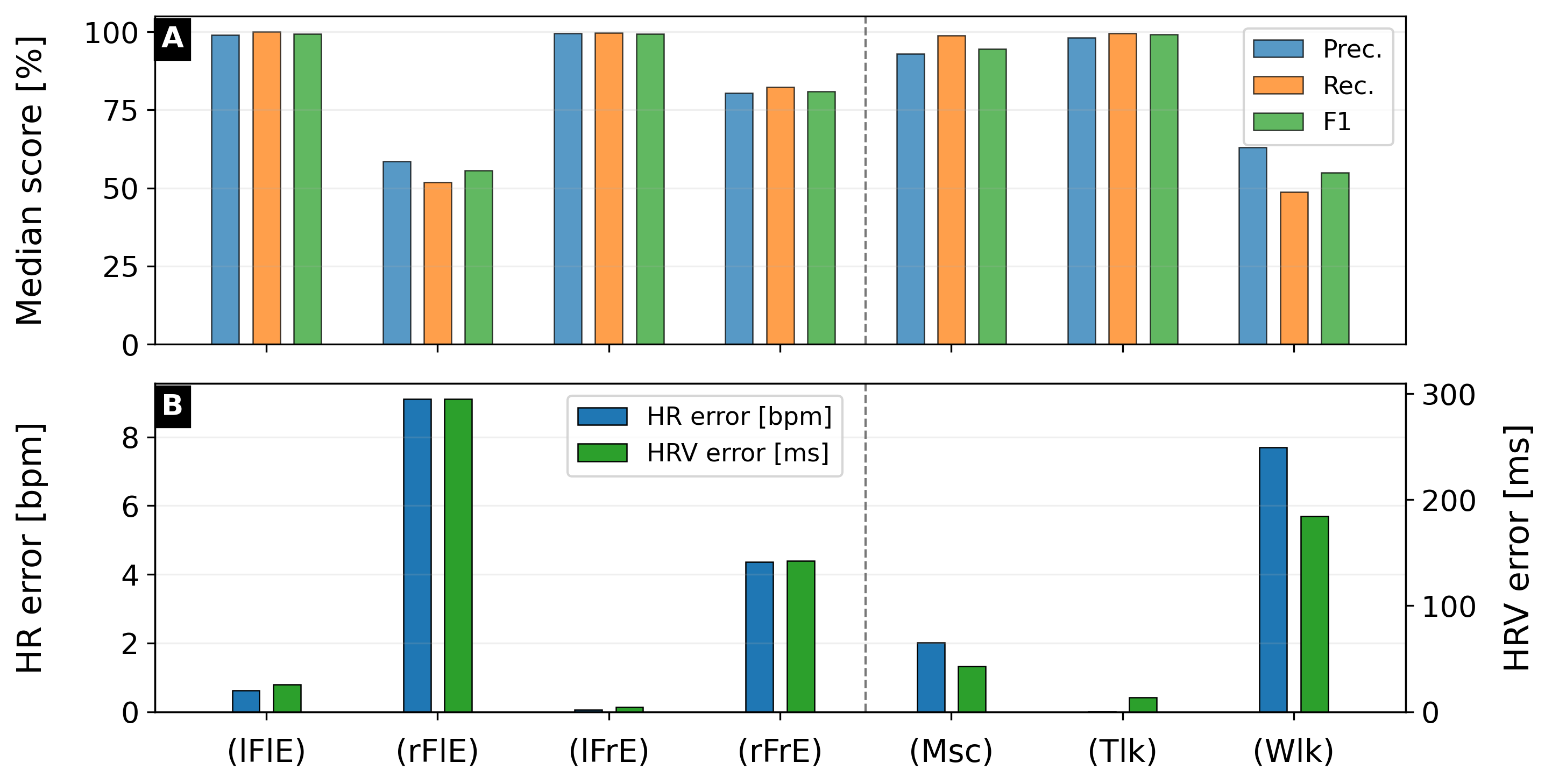}
    \caption{Median heartbeat detection performance and HR/HRV agreement errors across all configurations \& noise conditions. [A]: Precision (blue), recall (orange), F1 score (green). [B]: HR error (blue) and HRV (SDNN) error (green).}
    \label{fig:precision-recall-hr-hrv-error-median}
\end{figure}

\autoref{fig:precision-recall-hr-hrv-error-median} shows R peak detection performance and HR/HRV agreement.
Among stationary conditions, \textit{lFrE} achieved the highest overall performance, with a median R peak detection precision of 99.49\%, recall of 99.62\% and F1 score of 99.30\%.

Although \textit{lFlE} achieved a slightly higher recall (100.00\%) and F1 score (99.34\%), \textit{lFrE} yielded the lowest agreement errors overall (HR error: 0.05\,bpm; HRV error: 4.29\,ms).
In contrast, right-finger configurations showed substantially degraded performance, with \textit{rFrE} performing better (F1 = 80.95\%) than \textit{rFlE} (F1 = 55.59\%).
Under the disturbance conditions, \textit{Msc} (F1 = 94.54\%) maintained comparatively robust heartbeat detection performance compared to \textit{lFlE}, while \textit{Tlk} (F1 = 99.05\%) achieved performance nearly on par. Walking, however, substantially reduced performance, resulting in an F1-score of only 54.96\%.

\subsubsection{Heartbeat Metrics}

For every participant and every configuration, heartbeat rhythm metrics HR, SDNN, RMSSD and pNN50 were calculated, and correlated across configurations. The results are shown in the left half of \autoref{tab:combined-correlations}. 
The best agreement occurred for the cross-body left-finger configuration \textit{lFrE}, followed by \textit{lFlE}.
Right-finger configurations \textit{rFlE} and \textit{rFrE} yielded markedly reduced agreement across most heartbeat metrics, frequently without significant correlations. 
Under disturbance, \textit{Msc} showed low correlations, whereas \textit{Tlk} achieved unexpectedly high HR agreement even exceeding \textit{lFlE}, despite substantially reduced agreement for SDNN, RMSSD, and pNN50.

\input{resources/palmisciano_reduced_correlation_values.tex}
\input{resources/de_lucia_median_correlation_values.tex}

\begin{table}[h]
    \centering
    \caption{
    Correlations of heartbeat and waveform metrics between ear-ECG and reference across configurations and disturbance conditions. Bold values marked with asterisk (*) indicate significance after Bonferroni correction within their respective analysis family.}
    
    \label{tab:combined-correlations}

	\scriptsize

	\setlength{\tabcolsep}{2.5pt}

	\renewcommand{\arraystretch}{0.9}

	\resizebox{\linewidth}{!}{%
\begin{tabular}{c!{\vrule width 0.5pt}llll!{\vrule width 0.5pt}lllll}
        \toprule
		& \multicolumn{4}{c!{\vrule width 0.5pt}}{\textbf{Heartbeat Metrics}} & \multicolumn{5}{c}{\textbf{Waveform Metrics}} \\[2pt]
        \textbf{Conf.} & HR & SDNN & RMSSD & pNN50 & P\textsubscript{A} & QRS\textsubscript{A} & PR & QRS & QT \\
        \midrule

        \textbf{\textit{lFlE}} & \PalmiscianoOneHR & \PalmiscianoOneSDNN & \PalmiscianoOneRMSSD & \PalmiscianoOnePNN
            & \DeLuciaOnePWaveAmpl & \DeLuciaOneQRSAmpl & \DeLuciaOnePRInterval & \DeLuciaOneQRSDuration & \DeLuciaOneQTInterval \\

        \textbf{\textit{rFlE}} & \PalmiscianoTwoHR & \PalmiscianoTwoSDNN & \PalmiscianoTwoRMSSD & \PalmiscianoTwoPNN
            & \DeLuciaTwoPWaveAmpl & \DeLuciaTwoQRSAmpl & \DeLuciaTwoPRInterval & \DeLuciaTwoQRSDuration & \DeLuciaTwoQTInterval \\

        \textbf{\textit{lFrE}} & \PalmiscianoThreeHR & \PalmiscianoThreeSDNN & \PalmiscianoThreeRMSSD & \PalmiscianoThreePNN
            & \DeLuciaThreePWaveAmpl & \DeLuciaThreeQRSAmpl & \DeLuciaThreePRInterval & \DeLuciaThreeQRSDuration & \DeLuciaThreeQTInterval \\

        \textbf{\textit{rFrE}} & \PalmiscianoFourHR & \PalmiscianoFourSDNN & \PalmiscianoFourRMSSD & \PalmiscianoFourPNN
            & \DeLuciaFourPWaveAmpl & \DeLuciaFourQRSAmpl & \DeLuciaFourPRInterval & \DeLuciaFourQRSDuration & \DeLuciaFourQTInterval \\

        \midrule

        \textbf{\textit{Msc}} & \PalmiscianoFiveHR & \PalmiscianoFiveSDNN & \PalmiscianoFiveRMSSD & \PalmiscianoFivePNN
            & \DeLuciaFivePWaveAmpl & \DeLuciaFiveQRSAmpl & \DeLuciaFivePRInterval & \DeLuciaFiveQRSDuration & \DeLuciaFiveQTInterval \\

        \textbf{\textit{Tlk}} & \PalmiscianoSixHR & \PalmiscianoSixSDNN & \PalmiscianoSixRMSSD & \PalmiscianoSixPNN
            & \DeLuciaSixPWaveAmpl & \DeLuciaSixQRSAmpl & \DeLuciaSixPRInterval & \DeLuciaSixQRSDuration & \DeLuciaSixQTInterval \\

        \addlinespace[1pt]

        \textbf{\textit{Wlk}} & \PalmiscianoSevenHR & \PalmiscianoSevenSDNN & \PalmiscianoSevenRMSSD & \PalmiscianoSevenPNN
            & \DeLuciaSevenPWaveAmpl & \DeLuciaSevenQRSAmpl & \DeLuciaSevenPRInterval & \DeLuciaSevenQRSDuration & \DeLuciaSevenQTInterval \\

        \bottomrule
    \end{tabular}%
    }
\end{table}

\begin{figure}[t]
    \centering
    \includegraphics[width=\linewidth]{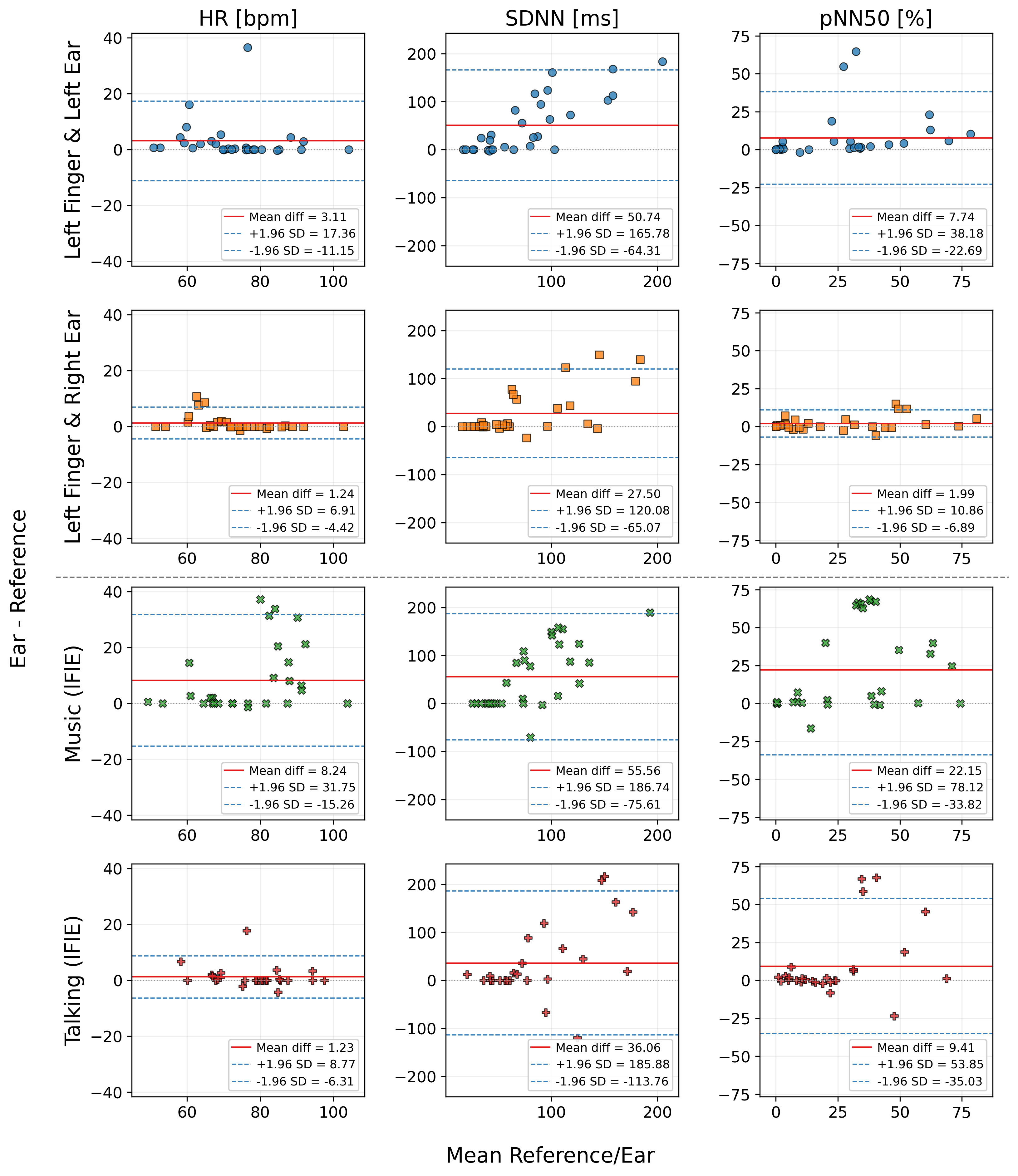}
    \caption{Bland-Altman plots of HR, SDNN \& pNN50 for left-finger configurations and stationary disturbance conditions.}
    \label{fig:palmisciano_bland_altman}
\end{figure}

Bland-Altman plots of selected metrics and configurations shown in \autoref{fig:palmisciano_bland_altman} support these observations: \textit{lFrE} exhibits the narrowest limits of agreement and smallest systematic bias, followed by \textit{lFlE}; disturbance conditions show larger spread.
Walking resulted in non-significant correlations across all metrics.

\subsubsection{ECG Waveform Metrics}
\begin{figure*}[t]
    \centering
    \includegraphics[width=\textwidth]{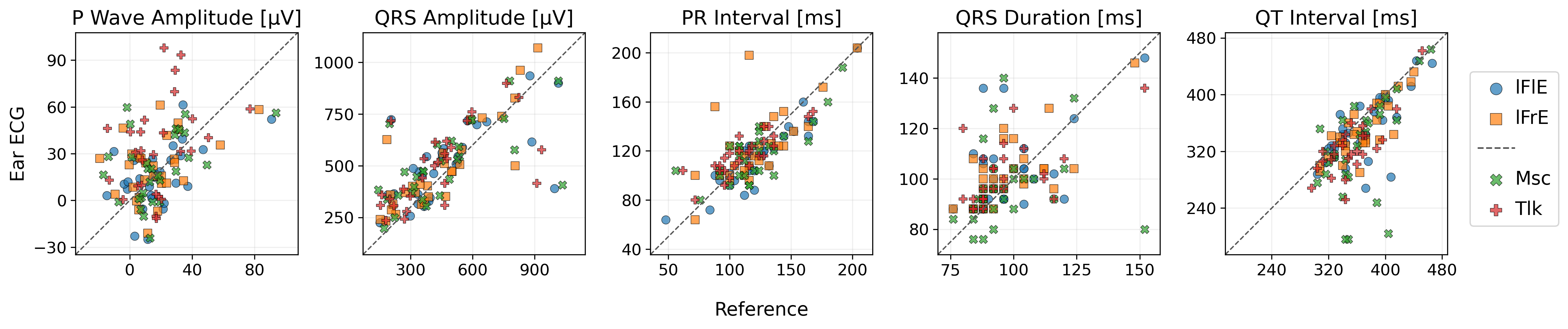}
    \caption{Scatter plots of waveform metrics for left-finger configurations and stationary disturbance conditions.}
    \label{fig:de-lucia-median-scatter}
\end{figure*}

Correlations of participant-level median ECG waveform metrics are shown on the right side of \autoref{tab:combined-correlations}. 
Overall, \textit{lFrE} produced the strongest agreement across most metrics, particularly QRS amplitude, QRS duration, and QT interval. In contrast, \textit{lFlE} yielded stronger correlations for PR interval.
Under disturbance, \textit{Tlk} maintained slightly stronger morphology agreement than \textit{Msc}.
Scatter plots (\autoref{fig:de-lucia-median-scatter}) for \textit{lFlE}, \textit{lFrE}, \textit{Msc} and \textit{Tlk} visually support these results.
Walking and right-finger conditions showed the lowest morphology agreement, consistent with altered waveforms in \autoref{fig:avg-waveform}, indicating unreliable fiducial point preservation.

\subsection{Finger-to-Ear ECG User Comfort}
\begin{figure}[h]
    \centering
    \includegraphics[width=\linewidth]{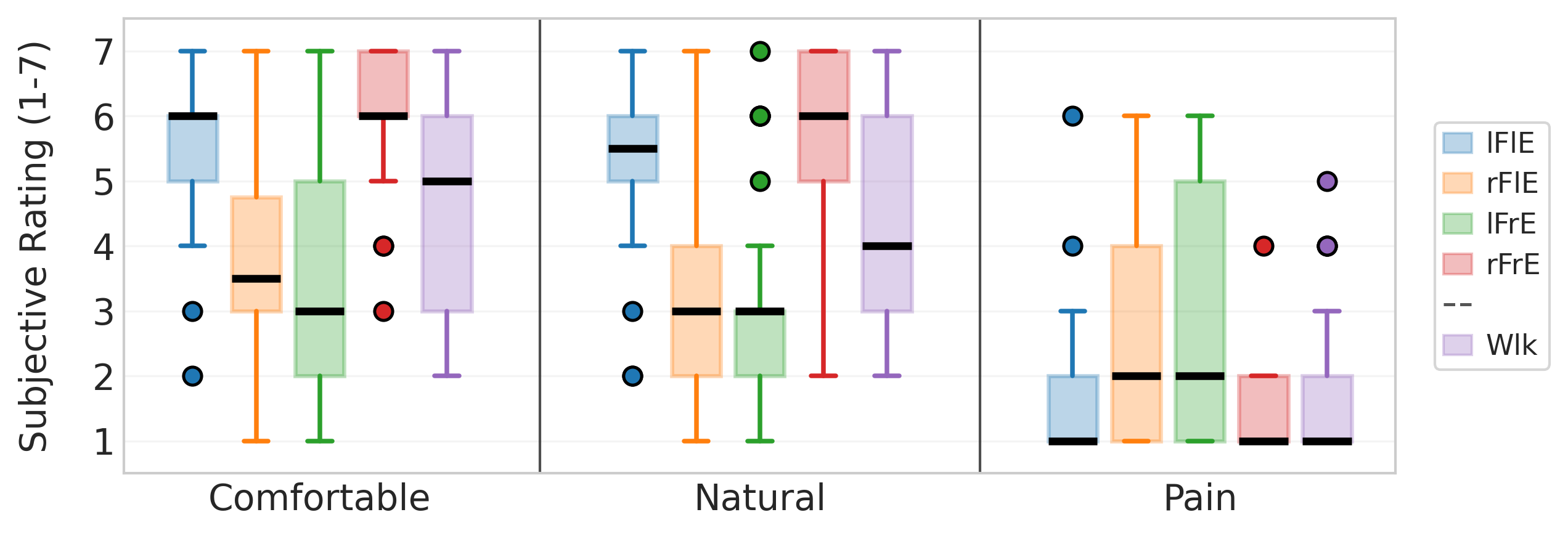}
    \caption{Box-Whisker plots of comfort, naturalness and pain assessments for all four configurations and while walking.}
    \label{fig:Box-Whisker}
\end{figure}

\autoref{fig:Box-Whisker} shows questionnaire ratings for perceived comfort, naturalness, and pain of all configurations and during walking. 
Same-sided configurations yielded high median ratings (\textit{lFlE}: comfort: 6, natural 5.5; \textit{rFrE}: comfort 6, natural 6). 
Performing a \textit{lFlE} ECG recording while walking resulted in slightly reduced comfort (median rating: 5), with increased variance. 
Cross-body configurations were rated more uncomfortable (medians 3–3.5), unnatural (median 3), and painful (median 2).
Friedman tests revealed significant differences across all dimensions ($p < 0.001$).

Kendall's $W$ effect sizes indicate \ComfortWInterpretation{} agreement for comfort ($W=\ComfortKendallsW$), \NaturalWInterpretation{} agreement for naturalness ($W=\NaturalKendallsW$), and \PainWInterpretation{} agreement for pain ($W=\PainKendallsW$). 
Post-hoc pairwise comparisons (Wilcoxon signed-rank tests with Bonferroni correction ($\alpha_{\text{adj}} = 0.005$)) revealed significant differences for same-sided versus cross-body contrasts in comfort ($p = \ComfortPosthocWorstP$) and naturalness ($p = \NaturalPosthocWorstP$); the pain contrast did not survive correction ($p = \PainPosthocWorstP$).

\section{Discussion}
This work introduced and systematically evaluated a wearable finger-to-ear ECG form factor. 
Signal quality strongly depended on electrode geometry: left-finger configurations consistently outperformed right-finger configurations across all metrics. 
This asymmetry likely arises because the right finger and head lie on the same side of the cardiac axis, resulting in a comparatively small potential difference~\cite{waller1889iv, de_lucia_-ear_2021}. 
The results further suggest that ventricular ECG features can be preserved more reliably than atrial features. 
QRS-related metrics and QT intervals showed stronger agreement than P wave amplitude, which aligns with previous ECG literature reporting that P wave detection is particularly challenging due to its low amplitude and susceptibility to noise and motion artefacts, especially in wearable recordings~\cite{hossain2019accurate, tecelao2018automated}. 
Overall, the results indicate that left-finger configurations enable partial ECG morphology assessment, although further clinical validation is required before diagnostic interpretation can be claimed.

Among the evaluated geometries, the cross-body configuration left finger \& right ear (\textit{lFrE}) achieved the best overall signal fidelity, but the same-side configuration left finger \& left ear (\textit{lFlE}) was perceived as substantially more comfortable and natural in the questionnaire results, highlighting a practical trade-off between signal quality and interaction comfort in finger-to-ear ECG acquisition. 

Compared with existing ear-ear or single-ear ECG literature, this work demonstrates substantial improvements in signal quality and robustness, showing that alternative ear-centred geometries can enhance waveform fidelity while maintaining a user-friendly interaction paradigm that avoids adhesive or otherwise inconvenient electrode placements.

\paragraph{Real-World Disturbances}
Common real-world disturbances such as talking or listening to music during ECG acquisition degraded signal quality but remained largely usable, with music playback causing slightly stronger degradation than talking. 
The reduced performance during talking is likely caused by small movement artefacts, whereas music playback presumably introduced electrical interference from the speaker. 
Future designs could mitigate these effects through improved filtering, shielding, and hardware separation between audio and ECG components.
In contrast, walking introduced severe motion artefacts that substantially degraded waveform morphology and rendered reliable analysis impractical. Addressing this limitation will require improved mechanical stabilisation and adaptive filtering approaches capable of separating motion-induced baseline wander from cardiac activity. 

Overall, the observed degradation patterns highlight the importance of evaluating ear-ECG systems under realistic everyday conditions rather than controlled laboratory settings alone.

\paragraph{Limitations}
Although the proposed approach demonstrated substantial benefits in signal quality and morphology preservation, the wearable finger-to-ear paradigm requires active finger contact, making passive long-term monitoring impossible.
Furthermore, the study population consisted only of young, healthy adults, limiting generalisability to broader populations.
The current prototype is not yet a fully self-contained true-wireless wearable system, and signal quality could likely be further improved through active grounding (DRL). 
Finally, although several metrics indicate preservation of ECG morphology, the clinical suitability for diagnostic interpretation has not yet been validated by medical experts.

\section{Conclusion}
This study evaluated a finger-to-ear ECG paradigm across four geometries and three disturbances. Results show that geometry determines signal quality: left-finger configurations achieved F1-scores >99\%, outperforming previous ear-ear methods without complex machine learning reconstruction. While cross-body configurations (left finger \& right ear) yielded peak fidelity, same-side contact (left finger \& left ear) provided the best trade-off between signal quality and user comfort. Ventricular features (QRS, QT) were well-preserved, though atrial morphology (P wave) remains noise-sensitive.


Compared with existing ear-ear and single-ear ECG approaches, the proposed system achieved substantially improved heartbeat detection and morphology preservation without relying on machine learning based reconstruction or denoising \cite{davies_deep-match_2024, santos_real-time_2025}.
Supported by our open-source hardware, these findings establish a new baseline for earable sensing. 

Future efforts must focus on adaptive motion compensation for gait and clinical validation for diagnostic applications.


\begin{acks}
This work was partially supported by funding from the program Core-Informatics of the Helmholtz Association (HGF), HEiKA funded project PACo and the DFG, German Research Foundation Project KD2School GRK2739.
\end{acks}

\bibliographystyle{ACM-Reference-Format}
\balance
\bibliography{references}

\end{document}

%% file: participant_summary.tex
\newcommand{\Participants}{30}
\newcommand{\AgeSummary}{$26.57 \pm 5.13$ years}
\newcommand{\SexSummary}{21 males and 9 females}
\newcommand{\HandednessSummary}{28 right-handed participants and 2 left-handed participants}

%% file: questionnaire_statistics.tex
\newcommand{\ComfortKendallsW}{0.498}
\newcommand{\ComfortWInterpretation}{moderate}

\newcommand{\NaturalKendallsW}{0.589}
\newcommand{\NaturalWInterpretation}{strong}

\newcommand{\PainKendallsW}{0.446}
\newcommand{\PainWInterpretation}{moderate}

\newcommand{\ComfortPosthocWorstP}{0.001}

\newcommand{\NaturalPosthocWorstP}{0.002}

\newcommand{\PainPosthocWorstP}{0.014}



%% file: resources/yarici_mean_cardiac_rhythm_correlation_values.tex
\newcommand{\YaricilFlEMeanCardiacRhythmPearson}{0.90}
\newcommand{\YaricirFlEMeanCardiacRhythmPearson}{-0.30}
\newcommand{\YaricilFrEMeanCardiacRhythmPearson}{0.92}
\newcommand{\YaricirFrEMeanCardiacRhythmPearson}{-0.27}
\newcommand{\YariciMscMeanCardiacRhythmPearson}{0.87}
\newcommand{\YariciTlkMeanCardiacRhythmPearson}{0.83}
\newcommand{\YariciWlkMeanCardiacRhythmPearson}{0.60}

%% file: resources/palmisciano_reduced_correlation_values.tex
\newcommand{\PalmiscianoOneHR}{\textbf{0.84}\textsuperscript{*}}
\newcommand{\PalmiscianoOneSDNN}{\textbf{0.64}\textsuperscript{*}}
\newcommand{\PalmiscianoOneRMSSD}{\textbf{0.63}\textsuperscript{*}}
\newcommand{\PalmiscianoOnePNN}{\textbf{0.81}\textsuperscript{*}}
\newcommand{\PalmiscianoTwoHR}{\textbf{0.62}\textsuperscript{*}}
\newcommand{\PalmiscianoTwoSDNN}{0.32}
\newcommand{\PalmiscianoTwoRMSSD}{0.21}
\newcommand{\PalmiscianoTwoPNN}{0.45}
\newcommand{\PalmiscianoThreeHR}{\textbf{0.97}\textsuperscript{*}}
\newcommand{\PalmiscianoThreeSDNN}{\textbf{0.72}\textsuperscript{*}}
\newcommand{\PalmiscianoThreeRMSSD}{\textbf{0.69}\textsuperscript{*}}
\newcommand{\PalmiscianoThreePNN}{\textbf{0.98}\textsuperscript{*}}
\newcommand{\PalmiscianoFourHR}{\textbf{0.67}\textsuperscript{*}}
\newcommand{\PalmiscianoFourSDNN}{0.25}
\newcommand{\PalmiscianoFourRMSSD}{0.16}
\newcommand{\PalmiscianoFourPNN}{\textbf{0.56}\textsuperscript{*}}
\newcommand{\PalmiscianoFiveHR}{\textbf{0.68}\textsuperscript{*}}
\newcommand{\PalmiscianoFiveSDNN}{0.20}
\newcommand{\PalmiscianoFiveRMSSD}{0.15}
\newcommand{\PalmiscianoFivePNN}{0.42}
\newcommand{\PalmiscianoSixHR}{\textbf{0.93}\textsuperscript{*}}
\newcommand{\PalmiscianoSixSDNN}{0.19}
\newcommand{\PalmiscianoSixRMSSD}{0.09}
\newcommand{\PalmiscianoSixPNN}{0.47}
\newcommand{\PalmiscianoSevenHR}{0.28}
\newcommand{\PalmiscianoSevenSDNN}{0.05}
\newcommand{\PalmiscianoSevenRMSSD}{0.07}
\newcommand{\PalmiscianoSevenPNN}{0.15}

%% file: resources/de_lucia_median_correlation_values.tex
\newcommand{\DeLuciaOnePWaveAmpl}{0.52}
\newcommand{\DeLuciaOneQRSAmpl}{\textbf{0.63}\textsuperscript{*}}
\newcommand{\DeLuciaOnePRInterval}{\textbf{0.92}\textsuperscript{*}}
\newcommand{\DeLuciaOneQRSDuration}{0.53}
\newcommand{\DeLuciaOneQTInterval}{\textbf{0.70}\textsuperscript{*}}
\newcommand{\DeLuciaTwoPWaveAmpl}{0.09}
\newcommand{\DeLuciaTwoQRSAmpl}{\textbf{0.61}\textsuperscript{*}}
\newcommand{\DeLuciaTwoPRInterval}{0.16}
\newcommand{\DeLuciaTwoQRSDuration}{-0.13}
\newcommand{\DeLuciaTwoQTInterval}{\textbf{0.57}\textsuperscript{*}}
\newcommand{\DeLuciaThreePWaveAmpl}{0.41}
\newcommand{\DeLuciaThreeQRSAmpl}{\textbf{0.85}\textsuperscript{*}}
\newcommand{\DeLuciaThreePRInterval}{\textbf{0.68}\textsuperscript{*}}
\newcommand{\DeLuciaThreeQRSDuration}{\textbf{0.70}\textsuperscript{*}}
\newcommand{\DeLuciaThreeQTInterval}{\textbf{0.81}\textsuperscript{*}}
\newcommand{\DeLuciaFourPWaveAmpl}{0.11}
\newcommand{\DeLuciaFourQRSAmpl}{0.21}
\newcommand{\DeLuciaFourPRInterval}{0.10}
\newcommand{\DeLuciaFourQRSDuration}{-0.25}
\newcommand{\DeLuciaFourQTInterval}{0.20}
\newcommand{\DeLuciaFivePWaveAmpl}{0.40}
\newcommand{\DeLuciaFiveQRSAmpl}{\textbf{0.64}\textsuperscript{*}}
\newcommand{\DeLuciaFivePRInterval}{\textbf{0.76}\textsuperscript{*}}
\newcommand{\DeLuciaFiveQRSDuration}{0.17}
\newcommand{\DeLuciaFiveQTInterval}{0.54}
\newcommand{\DeLuciaSixPWaveAmpl}{0.36}
\newcommand{\DeLuciaSixQRSAmpl}{\textbf{0.61}\textsuperscript{*}}
\newcommand{\DeLuciaSixPRInterval}{\textbf{0.85}\textsuperscript{*}}
\newcommand{\DeLuciaSixQRSDuration}{0.57}
\newcommand{\DeLuciaSixQTInterval}{\textbf{0.76}\textsuperscript{*}}
\newcommand{\DeLuciaSevenPWaveAmpl}{-0.03}
\newcommand{\DeLuciaSevenQRSAmpl}{0.38}
\newcommand{\DeLuciaSevenPRInterval}{-0.20}
\newcommand{\DeLuciaSevenQRSDuration}{0.16}
\newcommand{\DeLuciaSevenQTInterval}{0.02}